\documentclass[twocolumn,superscriptaddress,nofootinbib,notitlepage,longbibliography,aps,prd]{revtex4-2}
\pdfoutput=1

\usepackage{aas_macros}
\usepackage{graphicx}
\usepackage{amsmath}
\usepackage{amssymb}
\usepackage{amsfonts}
\usepackage[dvipsnames]{xcolor}
\usepackage[linktoc=none]{hyperref}
\usepackage[utf8]{inputenc}
\usepackage{comment}

\usepackage{lineno}

\definecolor{tab-blue}{RGB}{0, 107, 164}
\hypersetup{colorlinks=true,allcolors=tab-blue}

\newcommand{\INFN}{INFN - Sezione di Napoli, Complesso Universitario Monte Sant'Angelo, 80126 Napoli, Italy}
\newcommand{\SSM}{Scuola Superiore Meridionale, Via Mezzocannone 4, 80138 Napoli, Italy}
\newcommand{\GSSI}{Gran Sasso Science Institute (GSSI), Viale Francesco Crispi 7, 67100 L’Aquila, Italy}
\newcommand{\LNGS}{INFN-Laboratori Nazionali del Gran Sasso (LNGS), via G. Acitelli 22, 67100 Assergi (AQ), Italy}

\begin{document}

\title{Exploring memory-burdened primordial black holes \\ with ultra-high-energy cosmic-rays}

\author{Antonio Ambrosone}
\email{antonio.ambrosone@gssi.it}
\affiliation{\GSSI}
\affiliation{\LNGS}
\author{Marco Chianese}
\email{m.chianese@ssmeridionale.it}
\affiliation{\SSM}
\affiliation{\INFN}
\author{Carmelo Evoli}
\email{carmelo.evoli@gssi.it}
\affiliation{\GSSI}
\affiliation{\LNGS}

\begin{abstract}
Quantum backreaction effects may quench Hawking evaporation through a ``memory burden'', allowing primordial black holes (PBHs) with formation masses well below $10^{15}~\mathrm{g}$ to survive to the present and contribute to the dark matter. We show that ultra-high-energy cosmic rays (UHECRs) provide a powerful and previously unexplored probe of this scenario. We compute the proton and neutron emission from memory-burdened PBHs, including the Galactic-halo contribution and the extragalactic proton component, and confront it with the Pierre Auger Observatory proton spectrum and its EeV neutron limits from the Galactic plane. This yields new constraints on the PBH dark-matter fraction as a function of the PBH formation mass and the evaporation-suppression parameter $k$. For $k\gtrsim 3$ the non-observation of ultra-high-energy protons leads to bounds competitive with those from UHE gamma rays, while neutron limits remain comparable to high-energy neutrino constraints. 
Our results highlights the key role of multi-messenger astronomy in constraining beyond-the-standard-model scenarios.
\end{abstract}

\maketitle

\section{Introduction}

Ultra-High-Energy Cosmic Rays~(UHECRs) are the most energetic particles ever observed reaching energies of the order of $10^{20}\, \rm eV$~\cite{PierreAuger:2021hun, 2013ApJ...768L...1A}. Despite their sources remain yet to be discovered, UHECRs allow us to probe fundamental physics at energies not accessible by any other experiment on Earth~\cite{Aloisio:2025nts, Chianese:2021jke, PierreAuger:2022ubv, PierreAuger:2022jyk, Das:2023wtk, Berat:2022iea, Aloisio:2007bh, PierreAuger:2025jwt, PierreAuger:2023vql, PierreAuger:2021tog, Chianese:2026cfz}.  In this article, for the first time, we explore the constraints imposed by UHECRs on memory-burdened Primordial Black Holes~(PBHs). PBHs represent hypothetical black holes produced by gravitational overdensities in the early Universe~\cite{Zeldovich:1967lct, Hawking:1971ei, Carr:1974nx} and they lately sparkled the interest of the astroparticle community because of their particle production due to Hawking radiation~\cite{Hawking:1974rv}. Indeed, PBHs should be unstable and thermally emit particles depending on their mass~\cite{Arbey:2019mbc, Arbey:2021mbl}. Within the conventional scenario, only PBHs with masses above $10^{15}~{\rm g}$ survive longer than the current age of the Universe, thereby remaining viable Dark Matter (DM) candidates. Nevertheless, existing observational constraints strongly limit the possibility that PBHs constitute the entirety of the DM abundance, confining this scenario primarily to the asteroid-mass window $10^{17}~{\rm g}\lesssim M_{\rm PBH} \lesssim 10^{22}~{\rm g}$~\cite{Carr:2016drx, Green:2020jor, Carr:2020gox, Carr:2021bzv}. In contrast, lighter PBHs with $M_{\rm PBH} \lesssim 10^{15}~{\rm g}$, despite having evaporated by the present epoch, can still play an important phenomenological role, as their presence in the early Universe may have impacted several cosmological processes. For example, PBHs have been proposed to influence mechanisms responsible for generating the baryon asymmetry~\cite{Ambrosone:2021lsx,Fujita:2014hha, Hamada:2016jnq, Morrison:2018xla, Chen:2019etb, Perez-Gonzalez:2020vnz, Datta:2020bht, Hooper:2020otu, JyotiDas:2021shi, DeLuca:2021oer, Bernal:2022pue, Calabrese:2023key, Calabrese:2023bxz, Schmitz:2023pfy, Barman:2024slw, Gunn:2024xaq,sym11040511}, act as sources of gravitational waves~\cite{Papanikolaou:2020qtd, Domenech:2020ssp, Papanikolaou:2022chm, Ireland:2023avg, Domenech:2024wao,Belotsky:2018wph}, and contribute to DM production mechanisms~\cite{Bernal:2020kse, Gondolo:2020uqv, Bernal:2020ili, Bernal:2020bjf, Cheek:2021odj, Cheek:2021cfe, Samanta:2021mdm, Bernal:2021yyb, Bernal:2021bbv, Sandick:2021gew, Bernal:2022oha, Cheek:2022mmy, Gehrman:2023qjn, Bertuzzo:2024fns, Franciolini:2026fdv,Cai:2023ptf,Yuan:2023bvh,Belotsky:2014kca,Khlopov:2020vpx,Khlopov:2024nqp,Khlopov:2008qy}. In this framework, Refs.~\cite{Dvali:2018xpy, Dvali:2020wft, Dvali:2024hsb} have recently explored the back-reaction of particles on the quantum states of black holes, deducing that the evaporation could be quenched when a big fraction of the initial mass is lost by the black hole. This effect, usually referred to as {\it memory burden}, implies that light PBHs with masses $M \lesssim 10^{15}\, \rm g$ could still be evaporating, leading to important phenomenological signatures~\cite{Chianese:2025wrk, Chianese:2024rsn, Alexandre:2024nuo, Thoss:2024hsr, Haque:2024eyh, Balaji:2024hpu, Barman:2024iht, Bhaumik:2024qzd, Barman:2024ufm, Kohri:2024qpd, Jiang:2024aju, Zantedeschi:2024ram, Barker:2024mpz, Borah:2024bcr, Loc:2024qbz, Basumatary:2024uwo, Athron:2024fcj, Bandyopadhyay:2025ast, Calabrese:2025sfh, Boccia:2025hpm, Liu:2025vpz, Dvali:2025ktz, Montefalcone:2025akm, Dondarini:2025ktz,  Levy:2025lyj, Tseng:2025fjf, Chaudhuri:2025asm, Kitabayashi:2025iaq, Tan:2025vxp, Chaudhuri:2025rcs, Maity:2025ffa, Merchand:2025bzt}.
Given that the typical energy of the emitted particles is set by the PBH temperature, which in turn depends inversely on its mass, sufficiently light PBHs can radiate extremely energetic particles. In particular, for $M_{\rm PBH} \lesssim 10^4\,\mathrm{g}$, the characteristic emission energy reaches $\gtrsim 10^{18}\,\mathrm{eV}$. In this regime, UHECRs provide a unique and previously unexplored avenue to probe memory-burdened PBHs, opening a complementary window with respect to traditional gamma-ray and neutrino searches.
Specifically, we exploit the UHECR flux and composition measurements~\cite{PierreAuger:2021hun, PhysRevD.102.062005,AbdulHalim:20239/, AbdulHalim:2023Yd, PierreAuger:2024flk} as well as the upper limits on Ultra-High-Energy (UHE) neutrons~\cite{PierreAuger:2014tey} set by the Pierre Auger Observatory in order to constrain the abundance of PBHs in a range of energy not probed before.  The paper is organized as follows. In Sec.~\ref{sec:memory} we detail the Hawking evaporation of memory-burdened PBHs. In Sec.~\ref{sec:CR} we describe the computation of the UHECR flux from the galactic and extragalactic distributions of PBHs considered as a DM component. In Sec.~\ref{sec:stat} we discuss the data samples and the statistical analysis used to derive the constraints on parameter space of memory-burdened PBH, which are then reported in Sec.~\ref{sec:results}. Finally, in Sec.~\ref{sec:conclusions} we draw our conclusions.

\section{Evaporation of memory-burdened black holes}
\label{sec:memory}

Quantum field theory predicts that black holes are not perfectly stable objects but instead emit particles continuously. This phenomenon, known as Hawking evaporation, produces radiation that is nearly thermal, with a characteristic temperature set by the black hole mass:
\begin{equation}\label{eq:temperature}
T_{\rm H} = \frac{1}{8\pi G M_{\rm PBH}} \simeq 10^{9}\left(\frac{10^{4}~{\rm g}}{M_{\rm PBH}}\right){\rm GeV}\,,
\end{equation}
where $G$ is Newton’s constant and $M_{\rm PBH}$ denotes the PBH mass. This emission induces a gradual decrease of the PBH mass over time. This mass depletion is governed by
\begin{equation}\label{eq:mass-loss}
\frac{{\rm d}M_{\rm PBH}}{{\rm d}t}
= - \frac{\mathcal{G} g_{\rm SM}}{30720 \pi G^2 M_{\rm PBH}^2}\,,
\end{equation}
where $\mathcal{G}\simeq 3.8$ parametrizes the gray-body corrections arising from the scattering of particles in the curved spacetime surrounding the black hole~\cite{Page:1976wx}. The factor $g_{\rm SM}\simeq 102.6$ counts the effective number of relativistic Standard Model degrees of freedom accessible at high Hawking temperatures~\cite{Mazde:2022sdx}. Within the standard evaporation picture, this process proceeds uninterrupted until the PBH mass is entirely radiated away. Integrating the mass-loss equation~\eqref{eq:mass-loss} yields the total lifetime of a PBH:
\begin{equation}
\tau_{\rm PBH}
= \frac{10240 \pi G^2 M_{\rm PBH}^3}{\mathcal{G} g_{\rm SM}}
\simeq 4.4\times 10^{17}
\left(\frac{M_{\rm PBH}}{10^{15}~{\rm g}}\right)^3~{\rm s}\,.
\end{equation}
Consequently, PBHs lighter than approximately $10^{15}~{\rm g}$ are expected to have completely evaporated by the present epoch.

In the memory-burden scenario, instead, the evaporation process separates into two distinct regimes. At early times, the PBH evolves as in the conventional Hawking picture, with its mass decreasing according to the standard evaporation law given by Eq.~\eqref{eq:mass-loss}. This semiclassical regime persists until quantum information stored on the horizon becomes dynamically relevant. In the limit of an abrupt onset of the memory effects, this change of behavior takes place at
\begin{equation}
t_q = \tau_{\rm PBH}(1-q^3)\,,
\end{equation}
by which point the PBH retains only a fraction $q$ of its original mass,
\begin{equation}
M_{\rm PBH}^{\rm mb} = q M_{\rm PBH}\,.
\end{equation}
Beyond $t_q$, the rate of mass loss is suppressed relative to the Hawking prediction as
\begin{equation}
\label{eq:rate}
\frac{{\rm d}M_{\rm PBH}^{\rm mb}}{{\rm d}t}
= S(M_{\rm PBH})^{-k}
\frac{{\rm d}M_{\rm PBH}}{{\rm d}t}\,,
\end{equation}
where $k>0$ controls the strength of the suppression which depends on the black hole's entropy defined by the Bekenstein–Hawking formula $S(M_{\rm PBH}) = 4\pi G M_{\rm PBH}^2$. Integrating the modified evolution equation yields the PBH mass as a function of time during the memory-dominated regime:
\begin{equation}
M_{\rm PBH}^{\rm mb}(t)
= M_{\rm PBH}^{\rm mb}
\left[1-\Gamma_{\rm PBH}^{(k)}(t-t_q)\right]^{\frac{1}{3+2k}}\,,
\end{equation}
where the effective decay rate is equal to
\begin{equation}
\Gamma_{\rm PBH}^{(k)}
= \frac{\mathcal{G} g_{\rm SM}}{7680\pi}
2^k(3+2k) M_{\rm P}
\left(\frac{M_{\rm P}}{M_{\rm PBH}^{\rm mb}}\right)^{3+2k}\,,
\end{equation}
with $M_{\rm P} = 1/\sqrt{8\pi G}$ denoting the reduced Planck mass.

The suppression of evaporation substantially prolongs the PBH lifetime. The total evaporation time is therefore dominated by the memory-burdened phase and can be approximated as
\begin{equation}
\tau_{\rm PBH}^{(k)}
= t_q + \left(\Gamma_{\rm PBH}^{(k)}\right)^{-1}
\simeq \left(\Gamma_{\rm PBH}^{(k)}\right)^{-1}\,.
\end{equation}
This extended lifetime may exceed the standard Hawking evaporation time by many orders of magnitude, opening the possibility that PBHs with much smaller initial masses survive until today. Such long-lived remnants could thus constitute a non-negligible component of the present-day DM abundance.

Throughout this work we adopt a reference value $q=1/2$, motivated by the expectation that quantum memory effects become dynamically important once a black hole has radiated away roughly half of its initial mass. The exact numerical choice of $q$ does not affect the qualitative behavior of the evaporation process, but instead determines how the derived bounds map onto the PBH mass at formation.
Moreover, for the sake of simplicity, we model the onset of the memory-dominated regime as occurring instantaneously, treating the crossover from semiclassical evaporation to the suppressed phase as instantaneous. While this approximation captures the essential phenomenology relevant for our purposes, recent works~\cite{Dvali:2025ktz, Montefalcone:2025akm} indicate that a smooth or delayed transition may occur, thus modifying the PBH mass evolution over cosmological timescales. Importantly, the observational constraints considered in this study are mainly driven by present-day CR emission from the galactic halo. As a result, our bounds are largely insensitive to the detailed early-time behavior of the evaporation process. The limits presented here can therefore be straightforwardly reinterpreted within alternative frameworks featuring non-instantaneous suppression, provided the corresponding modification to the late-time evaporation rate is specified. A systematic treatment of such extended scenarios is left for future investigation.

\section{The cosmic-ray flux from primordial black holes}
\label{sec:CR}

We here detail the computation of the CR flux emitted by a population of memory-burdened PBHs, which accounts for a fraction $f_{\rm PBH}=\Omega_{\rm PBH}/\Omega_{\rm DM}$ of the total DM density of the Universe. We assume a monochromatic PBH mass spectrum within the range $10^{-1}~{\rm g} \leq M_{\rm PBH} \leq 10^7 ~{\rm g}$. 

In the presence of memory effects, the Hawking emission is globally attenuated according to Eq.~\eqref{eq:rate}. As a result, the primary emission rate of particles of species $i$ from a non-rotating, neutral PBH of mass $M_{\rm PBH}$ is expected to be 
\begin{equation}
\frac{{\rm d}^2N^{\rm mb}_{i}}{{\rm d}E{\rm d}t} = S(M_{\rm PBH})^{-k} \frac{{\rm d}^2N_{i}}{{\rm d}E {\rm d}t}\,,
\end{equation}
where the quantity
\begin{equation}
    \label{eq:pri_spec}
    \frac{{\rm d}^2N_i}{{\rm d}E{\rm d}t} = \frac{g_i}{2 \pi} \frac{\mathcal{F}_i(E,M_{\rm PBH})}{e^{E/T_{\rm H}} - (-1)^{2s_i}} \,,
\end{equation}
is the semi-classical emission rate with $g_i$ and $s_i$ denoting the internal degrees of freedom and the spin of the particles, respectively, and $\mathcal{F}_i(E,M_{\rm PBH})$ being the gray-body factor. The suppression factor of $S(M_{\rm PBH})^{-k}$ affects the normalization of the spectrum without altering its characteristic energy scale: the maximum of the distribution continues to track the Hawking temperature, as in the standard semi-classical description. The primary emission from PBHs covers all elementary particles with masses below the Hawking temperature and, for light PBHs, it includes the entire Standard Model particle content.\footnote{PBHs emit particles and antiparticles with equal probability. All spectra shown for species such as protons, neutrons, and neutrinos include the sum of particle and antiparticle contributions. The neutrino spectrum is computed following the appendix in Ref.~\cite{Dondarini:2025ktz} in order to avoid double counting of the primary emission.} Hadrons such as protons and neutrons are then produced by the evolution and hadronization of the primary particles. Hence, the secondary spectrum of protons and neutrons can be computed through the convolution of the primary spectra with the fragmentation functions~\cite{Bauer:2020jay}.
\begin{figure}[t!]
\centering
\includegraphics[width=0.92\linewidth]{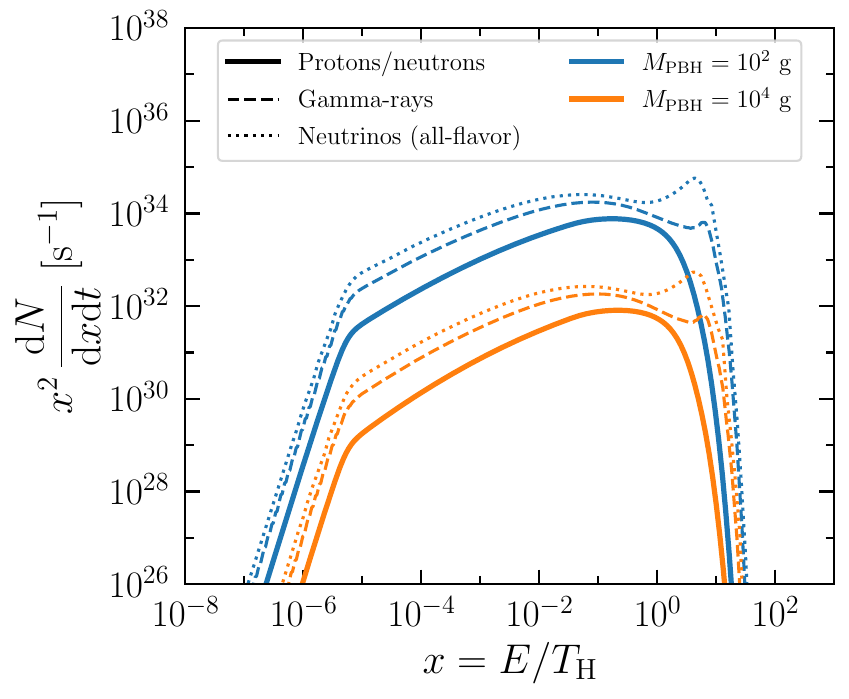}
    \caption{Proton/neutron (solid lines), gamma-ray (dashed lines) and neutrino (dotted lines) energy spectra from PBH evaporation, shown as functions of $x = E/T_{\rm H}$. The proton, neutron, and neutrino spectra include the sum of particle and antiparticle contributions. Results are displayed for two benchmark PBH masses: $M_{\rm PBH} = 10^2\,\mathrm{g}$ (blue curves) and $M_{\rm PBH} = 10^4\,\mathrm{g}$ (orange curves).}
    \label{fig:spectra}
\end{figure}

In this analysis, we use the publicly-available codes \texttt{BlackHawk}~\cite{Arbey:2019mbc, Arbey:2021mbl} and \texttt{HDMSpectra}~\cite{Bauer:2020jay} which provides tables for the primary emission spectra of PBHs and the fragmentation functions, respectively. The latter, however, only provides the fragmentation functions of the particle $i$ into protons, $\mathcal{D}_{i \to p}$. We therefore compute the spectrum of neutrons by assuming an approximate isopsin invariance between the up and down quarks, which implies $\mathcal{D}_{i \to p} \simeq \mathcal{D}_{i \to n}$~\cite{ParticleDataGroup:2024cfk}. We also emphasize that the relative abundances of the emitted species from the PBH evaporation only scale with their internal degrees of freedom (see Eq.~\eqref{eq:pri_spec}). In Fig.~\ref{fig:spectra}, we show the proton and neutron spectra from the evaporation of a single PBH, compared to the corresponding gamma-ray and neutrino spectra, as functions of the dimensionless variable $x = E/T_{\rm H}$, for representative benchmark values of $M_{\rm PBH}$. In general, all spectra peak at energies $E \sim T_{\rm H}$. Gamma-rays and neutrinos display fluxes larger by a factor of a few with respect to protons and neutrons, as they are produced both as primary particles and as secondary products of hadronization processes.

\begin{figure}[t!]
    \centering
    \includegraphics[width=0.92\linewidth]{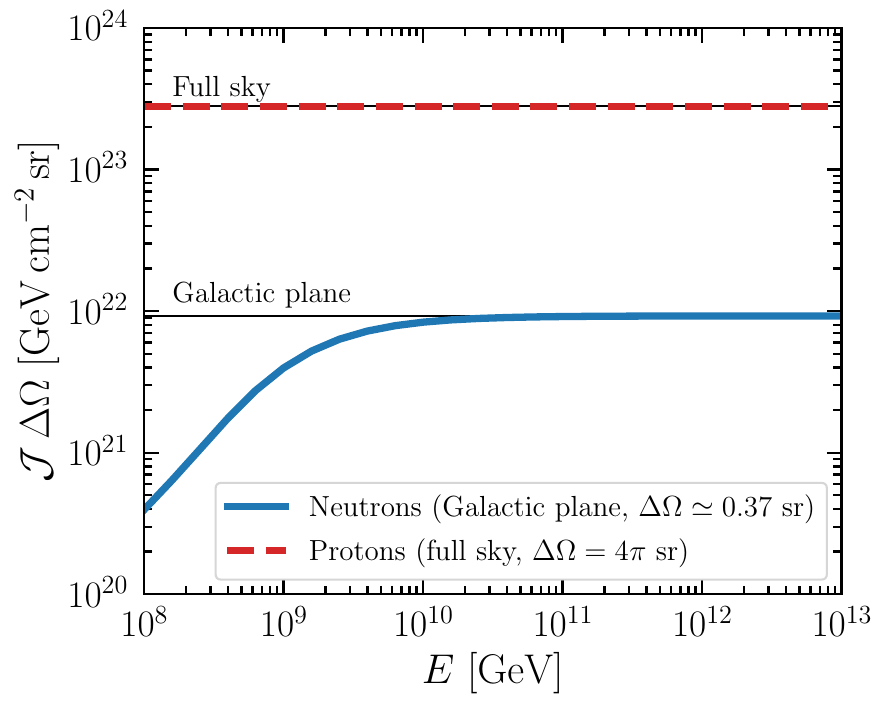}
    \caption{The angle-integrated J-factor as a function of the energy for the neutrons from the galactic plane~(blue line), defined as the region $0^{\circ}<l<360^{\circ}$ and $-1.7^{\circ}<b<{1.7^{\circ}}$ in galactic coordinates, and for the protons from the full sky (dashed red line).}
    \label{fig:Jfactor}
\end{figure}

In order to evaluate the CR flux from the whole population of PBHs, we take into account both the  galactic and extragalactic contributions. The angle-averaged galactic contribution from a region in the sky $\Delta \Omega$ can be calculated as
\begin{equation}\label{eq:galactic}
    \frac{{\rm d}^2\phi^{\rm gal}_{\rm CR}}{{\rm d}E {\rm d}\Omega} = \frac{f_{\rm PBH}}{4\pi M^{\rm mb}_{\rm PBH}}  \frac{{\rm d}^2 N^{\rm mb}_{\rm CR}}{{\rm d}E {\rm d}t} \mathcal{J}(E,\Delta \Omega)\,,
\end{equation}
where the quantity
\begin{equation}
    \mathcal{J}(E, \Omega) = \int_{\Delta \Omega} \frac{{\rm d}\Omega}{\Delta \Omega} \int_{0}^{+\infty} {\rm d}s \, \rho_{\rm PBH}(r) e^{-s/L_{\rm CR}(E)}\,,
\end{equation}
is the J-factor which depends on the PBH density distribution within the Galaxy, $\rho_{\rm PBH}(r)$, and takes into account the CR attenuation through the exponential factor. Following Refs.~\cite{Chianese:2024rsn, Chianese:2025wrk}, we consider the Navarro-Frank-White profile~\cite{Navarro:1995iw}  for the PBH galactic distribution, namely
\begin{equation}
    \rho_{\rm PBH}(r) = \frac{\rho_s}{{r}/{r_s}\left(1+{r}/{r_s}\right)^2}\, ,
\end{equation}
where $r_s = 25\, \rm kpc$ and $\rho_s = 0.23\, \rm GeV\, \rm cm^{-3}$~\cite{Benito:2020lgu}, and $r = (s^2+R_{\odot}^2 -2sR_{\odot} \cos b \cos l)^{1/2}$ is the radial distance from the galactic center with $R_{\odot} \simeq 8.178 \, \rm kpc$ being the Sun distance from the galactic center~\cite{Gravity:2019nxk}. In Eq.~\eqref{eq:galactic}, the quantity $L_{\rm CR}$ represents the decay length for cosmic rays, which is infinity for protons being stable particles, while for neutrons it is $L_n (E) \simeq 8.5 \left(E / {\rm EeV}\right)\,{\rm kpc} $~\cite{PierreAuger:2014tey}.
In the energy range relevant for this work, Galactic proton energy losses can be neglected because their interaction length is longer than typical galactic distances, in agreement with Ref.~\cite{Das:2023wtk}.

\begin{figure*}[t!]
    \centering
    \includegraphics[width=0.49\linewidth]{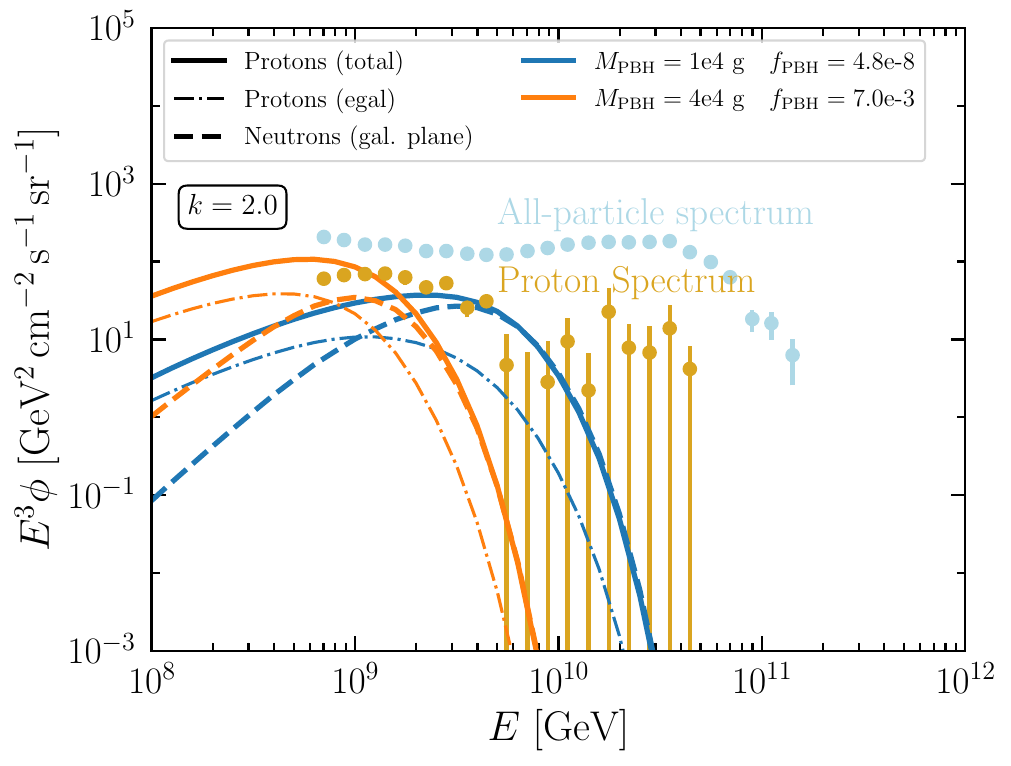}
     \includegraphics[width=0.49\linewidth]{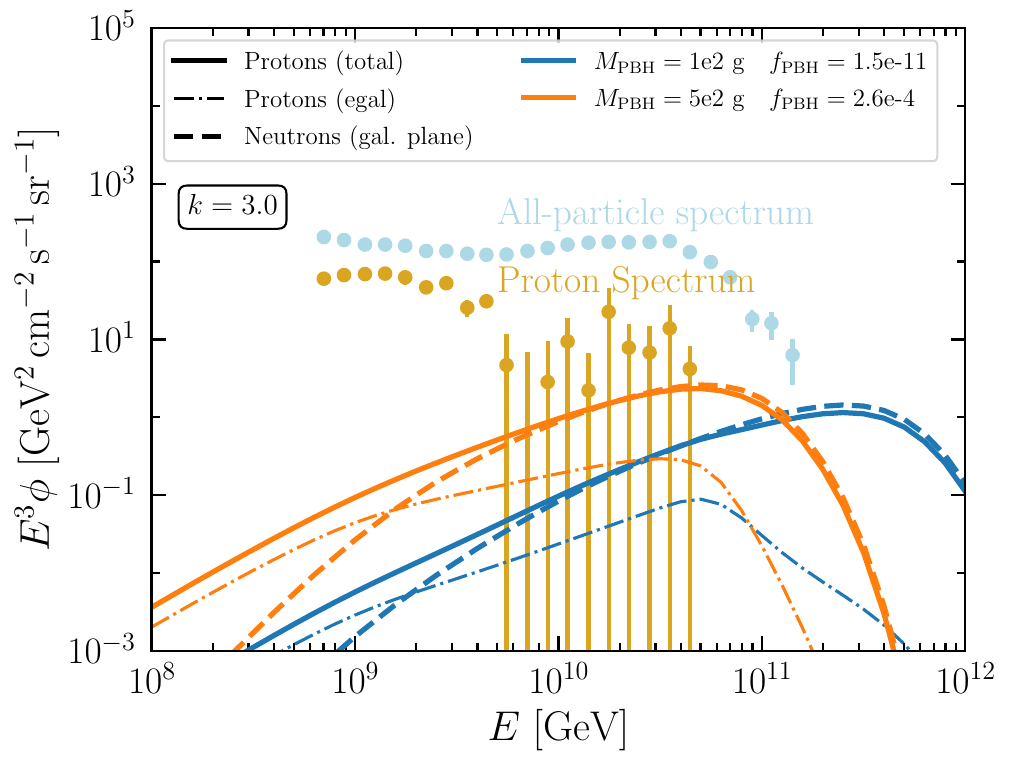}
    \caption{The Auger All-particle~\cite{PierreAuger:2021hun} and proton~\cite{AbdulHalim:2023Yd,AbdulHalim:20239/} spectra compared with the proton and neutron fluxes from PBHs fixing $f_{\rm PBH}$ at the 95\% CL upper limit with protons. Left: memory-burden scenario with $k=2$, for $M_{\rm PBH} = 10^4\, \rm g$ and $M_{\rm PBH} = 4 \times 10^{4}\, \rm g$. Right: memory-burden scenario with $k=3$, for $M_{\rm PBH} = 100\, \rm g$ and $M_{\rm PBH} = 500\, \rm g$. }
    \label{fig:flux_proton}
\end{figure*}

Figure~\ref{fig:Jfactor} reports the angle-integrated J-factor as a function of energy for the neutrons propagating in the galactic plane ($0^{\circ}<l<360^{\circ}$ and $-1.7^{\circ}<b<{1.7^{\circ}}$) and for the protons emitted from the full galactic volume. Neutrons exhibit an energy-increasing J-factor because of their decay, reaching about $4\%$ of the total galactic plane contribution at $E = 10^{8}\,\mathrm{GeV}$. In comparison, the galactic plane accounts for $\sim 3.7\%$ of the total J-factor of the Galaxy. However, due to the different sky coverage $\Delta \Omega$, the angle-averaged J-factor of the galactic plane results to be slightly larger than that of the full sky, the latter being equal to $2.2 \times 10^{22}~{\rm GeV\,cm^{-2}}$, in agreement with Refs.~\cite{Chianese:2024rsn, Chianese:2025wrk}. This is crucial because the neutron and proton data samples provide the integrated and the differential fluxes, respectively, implying a similar constraining power.

We account for the extragalactic protons, by solving their Boltzmann equation with the generation energy technique as explained in Refs.~\cite{Cermenati:2025ogl,Berezinsky:2002nc}
\begin{equation}\label{Eq:extragalactic_protons}
\begin{split}
    \frac{{\rm d}^2\phi_p^{\rm egal}}{{\rm d}E {\rm d}\Omega} = &  \frac{\Omega_{\rm DM}\rho_{\rm cr} f_{\rm PBH}}{4\pi M^{\rm mb}_{\rm PBH}}  \int_{0}^{z_{\rm max}} \frac{{\rm d}z}{H(z)} \times \\ & \qquad \; \left(
    \frac{1}{1+z}\frac{{\rm d}E_g}{{\rm d}E} \left.\frac{{\rm d}^2 N^{\rm mb}_{p}}{{\rm d}E {\rm d}t}\right|_{E=E_g(E,z)} \right)\, ,
\end{split}
\end{equation}
where $H(z) = H_0 \sqrt{\Omega_m (1+z)^3 + \Omega_{\Lambda}}$, with  $\Omega_m = 0.31$, $\Omega_{\Lambda} = 0.69$, $H_0 = 67.74\, \rm km\, \rm s^{-1}\, \rm Mpc^{-1}$ being the $\Lambda$CDM cosmological parameters reported by the Planck Collaboration and $\rho_{\rm cr} = 4.8 \times 10^{-6} \rm GeV\, \rm cm^{-3} $ being the critical density of the Universe~\cite{Planck:2018vyg}.  The expression is usually referred to as \textit{universal spectrum}, where $E_g(E,z)$ denotes exactly the \textit{generation energy}, that is the energy that a proton must have at redshift $z$ in order to be observed with energy $E$ today ($z=0$)~\cite{Cermenati:2025ogl}. 

This quantity consistently encodes all relevant energy-loss processes during propagation, including adiabatic losses from cosmic expansion and interactions with background photons, namely electron--positron pair and photopion productions. These interactions are computed considering only the Cosmic Microwave Background~(CMB) radiation, since the impact of the Extragalactic Background Light~(EBL) is negligible~\cite{Berezinsky:2002nc}. We implement pair-production and photopion losses following Ref.~\cite{Blumenthal:1970nn} and Ref.~\cite{Kelner:2008ke}, respectively.

The upper integration limit $z_{\rm max}$ corresponds to the maximal source redshift, which is set to $z_{\rm max}=5$, though we have verified that the contribution from higher values is negligible. We further assume that $f_{\rm PBH}$ is constant over cosmological timescales, thus fixing it to the same value as for the galactic component.

We neglect the direct extragalactic neutron component at Earth, since neutrons emitted by PBHs decay over distances much shorter than cosmological ones. For instance, at \(10^{21}\,\mathrm{eV}\) the neutron decay length is only \(\sim 8.5\,\mathrm{Mpc}\). Nevertheless, since neutron beta decay produces protons, we account for this contribution in the instantaneous-decay approximation by assuming that the proton injection term in Eq.~\ref{Eq:extragalactic_protons} is twice the direct proton injection from PBHs.


In Fig.~\ref{fig:flux_proton}, we compare the all-particle~\cite{PierreAuger:2021hun} and the proton~\cite{AbdulHalim:2023Yd, AbdulHalim:20239/} fluxes measured by the Pierre Auger Observatory with the corresponding spectra predicted from PBHs in case of some benchmark scenarios. The values of $f_{\rm PBH}$ are fixed to the $95\%$ CL upper limits derived from the proton analysis discussed in the next sections. The solid and dot-dashed lines correspond to the total (galactic and extragalactic) and the extragalactic proton fluxes, respectively. Furthermore, the dashed lines show the angle-averaged neutron fluxes emitted by the memory-burdened PBHs from the galactic plane. Interestingly, we find that its magnitude at high energies, where the neutron decays are negligible, is the same as that of the diffuse proton flux, despite the smaller J-factor as shown in Fig.~\ref{fig:Jfactor}.

\section{Statistical analysis}
\label{sec:stat}

\subsection{Neutrons}
The Pierre Auger Observatory has reported strong limits on the UHE neutrons from the galactic plane~\cite{PierreAuger:2014tey} (defined as the region with $0^{\circ}<l<360^{\circ}$ and $-1.7^{\circ}<b<{1.7^{\circ}}$) with an integrated upper limit above $1\, \rm EeV$ of $\Phi_n= 0.077\, \rm km^{-2}\, \rm yr^{-1}$ at 95\% Confidence Level~(CL). This indirectly constrains the neutrons emitted by memory-burdened PBHs according to the relation
\begin{equation}
    \int_{1\, \rm EeV}^{+\infty} {\rm d}E \int_{\Delta \Omega}{\rm d}\Omega \frac{{\rm d}^2\phi^{\rm gal}_n}{{\rm d}E {\rm d}\Omega} \lesssim \Phi_n\, .
\end{equation}
Here, the angular integral is evaluated over the galactic plane and the J-factor is given by the solid blue line in Fig.~\ref{fig:Jfactor}.

\begin{figure*}[t!]
    \centering
    \includegraphics[width=0.95\linewidth]{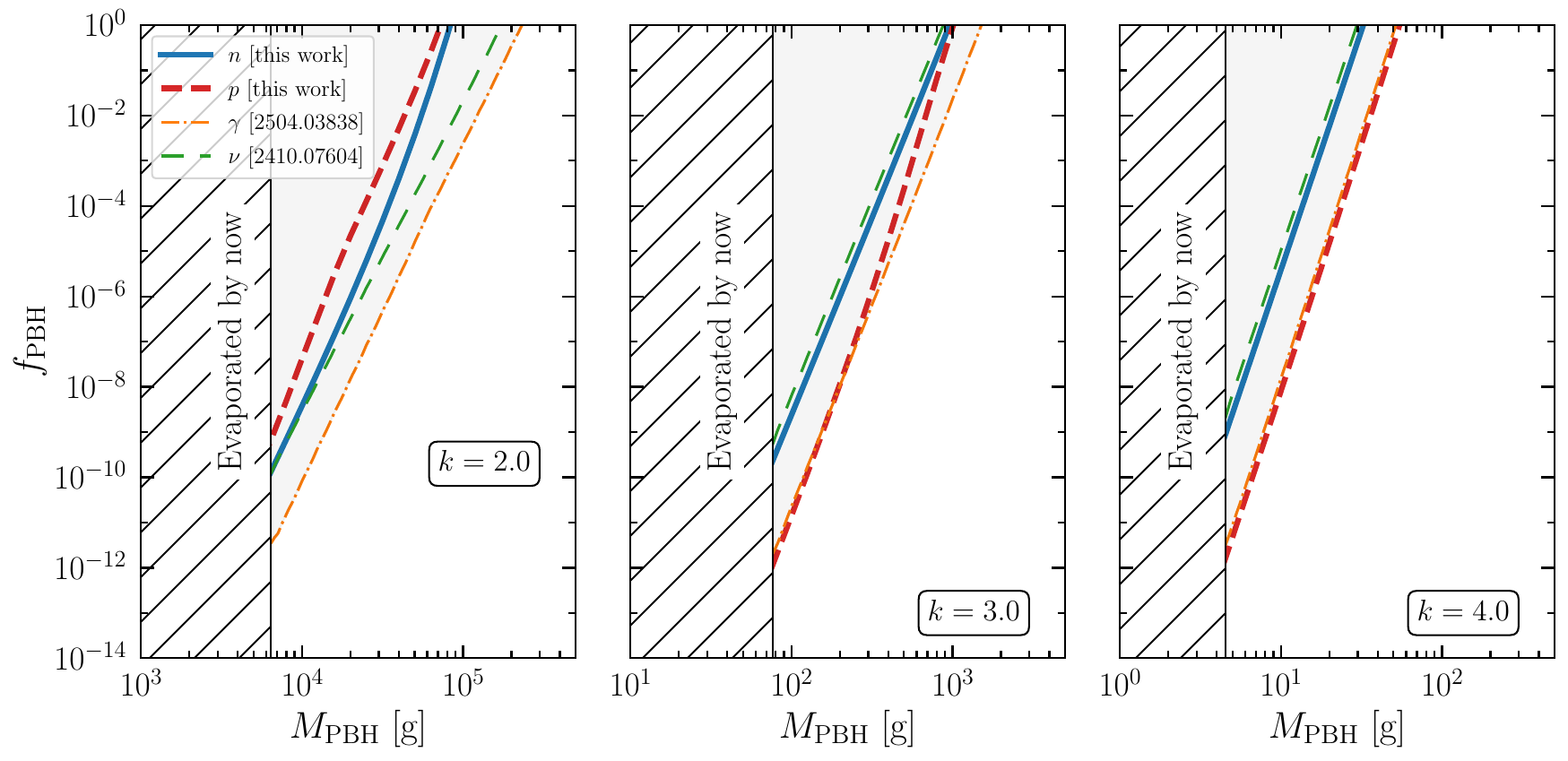}
    \caption{Constraints at 95\% CL on the fraction $f_{\rm PBH}$ as a function of the PBH mass at formation, in case of two scenarios for the memory-burden effect with $k = 2$ (left panel), $k = 3$ (middle panel)  and $k = 4$ (right panel). The solid blue and dashed red lines represent the bounds derived in the present work from UHE neutrons and protons, respectively. The dot-dashed orange and long-dashed green lines corresponds to the previous upper limits placed by UHE gamma-rays~\cite{Chianese:2025wrk} and neutrinos~\cite{Chianese:2024rsn}, respectively (see also Refs.~\cite{Alexandre:2024nuo, Thoss:2024hsr, Zantedeschi:2024ram, Boccia:2025hpm, Dondarini:2025ktz}). hatched region indicates PBHs that have fully evaporated over cosmological times.}
    \label{fig:limits_fpbh}
\end{figure*}

\subsection{Protons}
The measured UHECR flux extends from $\sim 7\times 10^{17}\, \rm eV$ up to $\sim 1.5 \times 10^{20}$~\cite{PierreAuger:2021hun}. The mass composition suggests a light mass composition up to $\sim 10^{19}\, \rm eV$, with a trend of a heavier composition required to fit the data at higher energies~\cite{PhysRevD.102.062005,AbdulHalim:20239/,AbdulHalim:2023Yd}.
In this work, to derive conservative proton-based constraints on PBHs, we adopt the proton spectrum~$(\phi^{\rm Auger}_{p}(E))$ inferred from the spectrum and composition measurements  by Auger using the \texttt{Sibyll 2.3d} interaction model~\cite{AbdulHalim:2023Yd}.
At energies above $5\times 10^{18}\, \rm eV$, where Auger reports only upper limits on the proton fraction, we impose that the proton spectrum does not exceed 
10\% of the all-particle spectrum~$(\phi^{\rm Auger}_{\rm all \, particle}(E))$. This choice is consistent with the proton fraction inferred in Refs.~\cite{AbdulHalim:2023Yd, AbdulHalim:20239/} and with the composition analysis~\cite{PierreAuger:2024flk}. Therefore, we define the chi-square difference as
\begin{equation}
    \Delta\chi^2_p \left(f_{\rm PBH}; \vec{\theta}_{\rm PBH} \right) = \sum_i\left\{
    \begin{array}{l l}
    \left(\frac{\mu_i - d_i}{\sigma_i}\right)^2 & \mu_i > d_i \\ 0 &\mu_i \leq d_i 
    \end{array}\right. \,,
\end{equation}
where the data $d_i$ represent the inferred best-fit flux
\begin{equation}
    d_i = \left\{\begin{array}{l l} 
    \phi^{\rm Auger}_{p}(E_i) & E_i \leq 5 \times 10^{18}~{\rm eV} \\
    0 & E_i > 5 \times 10^{18}~{\rm eV}
    \end{array} \right. \,,
\end{equation}
while $\sigma_i$ are the uncertainties on the measured flux, which is conservatively set equal to $0.1\,\phi^{\rm Auger}_{\rm all \, particle}(E)$ for $E \ge 5\times 10^{18}\, \rm eV$. 
\begin{equation}
    \mu_i \left(f_{\rm PBH}; \vec{\theta}_{\rm PBH} \right) = \frac{{\rm d}^2\phi^{\rm gal}_{p}}{{\rm d}E{\rm d}\Omega}(E_i) + \frac{{\rm d}^2\phi^{\rm egal}_{p}}{{\rm d}E{\rm d}\Omega}(E_i)\,,
\end{equation}
which is a function of the PBH abundance $f_{\rm PBH}$ and $\vec{\theta}_{\rm PBH} = \left(M_{\rm PBH},\,k\right)$.
The angle-averaged galactic proton flux is computed over the entire solid angle, {\it i.e.} considering the J-factor given by the dashed red line in Fig.~\ref{fig:Jfactor}. Thus, for each combination of $M_{\rm PBH}$ and $k$, we determine the maximum allowed value for $f_{\rm PBH}$ following the Wilks' theorem under the assumption of a single degree of freedom~\cite{Cowan:2010js}.

\section{Results and Discussion}
\label{sec:results}

Fig.~\ref{fig:limits_fpbh} reports the constraints on $f_{\rm PBH}$ in terms of $M_{\rm PBH}$, for the memory-burden scenarios with $k=2$~(left plot), $k=3$~(center plot) and $k=4$~(right plot). The results are also compared with the UHE gamma-rays and high-energy neutrino constraints obtained in Refs.~\cite{Chianese:2025wrk} and~\cite{Chianese:2024rsn}, respectively. These works adopt the same DM halo and likelihood analysis allowing a direct comparison between constraints with different messengers. However, for other constraints see Refs.~\cite{Alexandre:2024nuo, Thoss:2024hsr, Zantedeschi:2024ram, Boccia:2025hpm, Dondarini:2025ktz}. 

The hatched region denotes the parameter space in which PBHs fully evaporate over cosmological timescales, following the convention of Ref.~\cite{Thoss:2024hsr}.  However, since the constraints derived in this work are based on present-day UHECR observations, they apply only to PBHs that survive until today. Therefore, the hatched region is not probed by our analysis, although it may still be relevant for cosmological populations of PBHs that evaporated before the present epoch, as emphasized in Ref.~\cite{Chaudhuri:2025asm}.

Interestingly, for $k = 2$ neutrons impose stronger constraints than protons, despite probing only a limited region of the Galaxy.
For this value of $k$, UHE gamma rays provide the strongest constraints on the PBH abudance as shown in Refs.~\cite{Chianese:2025wrk,Thoss:2024hsr}.

By contrast, for $k = 3$  and $k = 4$ protons provide tighter constraints---by $\sim 1$--$2$ orders of magnitude---compared to neutrons. The origin of this difference lies in the energy dependence of the CR flux from PBHs: for $k = 2$ the emission peaks at $10^{9}$--$10^{10}\,\mathrm{GeV}$, where the measured proton flux is relatively high, leading to weaker proton constraints. Conversely, for $k \ge 3$ the PBH-induced flux is concentrated at higher energies, where the absence of observed protons severely restricts the allowed PBH emission. This is clearly shown in Fig.~\ref{fig:flux_proton}.

Interestingly, within this range of $k$, the bounds derived from protons are comparable in strength to those obtained from UHE gamma rays in Ref.~\cite{Chianese:2025wrk}. This result is non-trivial, given that gamma-ray fluxes are expected to be systematically larger than the proton ones, as discussed in Sec.~\ref{sec:CR}.
The origin of this behavior lies in the dependence on $k$: for larger values of this parameter, the relevant PBH mass window shifts towards lower masses, corresponding to higher particle emission energies. As a consequence, differential flux limits derived from the non-observation of ultra-high-energy protons become increasingly competitive with the integral constraints obtained from gamma rays. 
Furthermore, as illustrated in the central panel of Fig.~\ref{fig:limits_fpbh}, proton limits degrade more rapidly with increasing initial PBH mass. This is expected, since larger PBH masses correspond to spectra that peak at lower energies, where the sensitivity of proton measurements becomes progressively weaker. 
In fact, smaller values of $k$ would produce proton and neutron spectra peaking at lower energies, thereby leading to weaker CR constraints. In this regime, other messengers such as gamma-rays and neutrinos provide significantly stronger limits~\cite{Chianese:2025wrk, Chianese:2024rsn,Alexandre:2024nuo, Thoss:2024hsr, Zantedeschi:2024ram, Boccia:2025hpm, Dondarini:2025ktz}.

\begin{figure*}[t!]
    \centering
    \includegraphics[width=\linewidth]{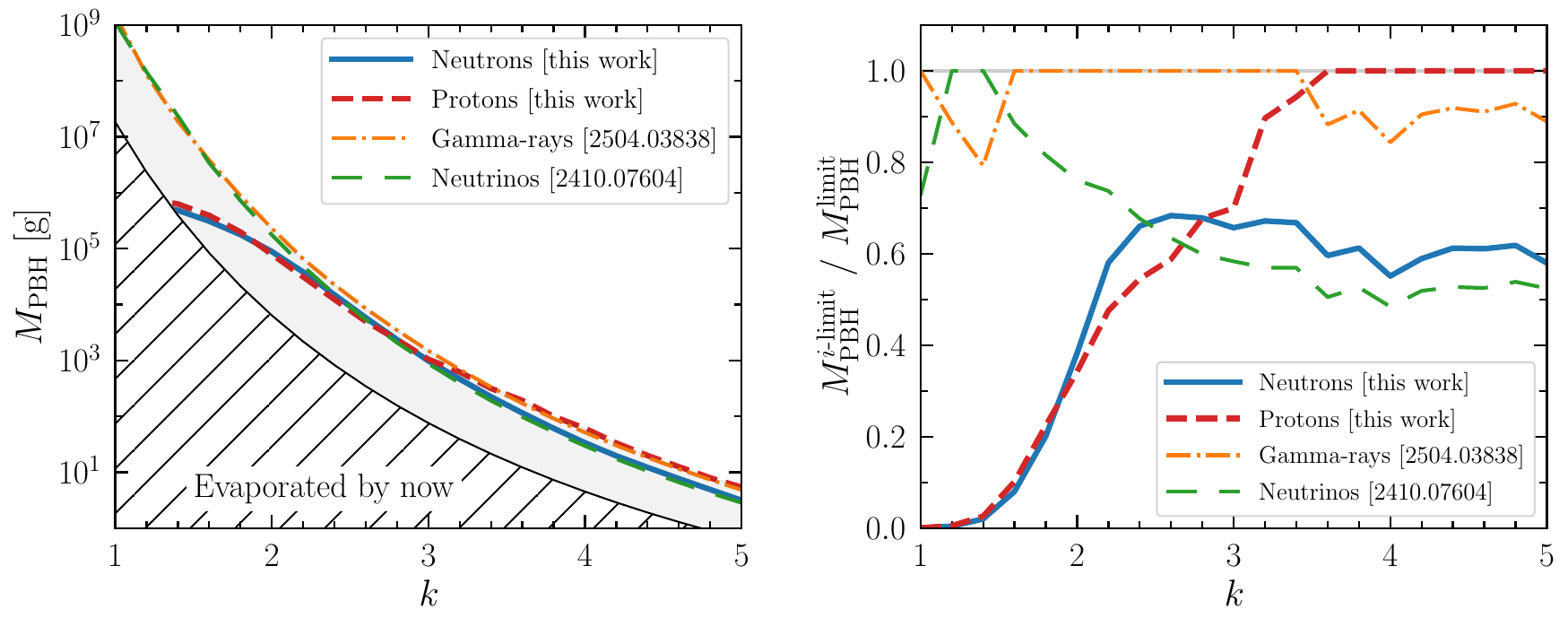}
    \caption{Constraints on memory-burdened PBHs as viable DM candidates. Left: the different lines refer to limits placed at 95\% CL in the $k$--$M_{\rm PBH}$ plane with $f_{\rm PBH} = 1$ according to UHE protons (solid blue line), UHE neutrons (dashed red line), UHE gamma-rays (dot-dashed orange line)~\cite{Chianese:2025wrk} and neutrinos (long-dashed green line)~\cite{Chianese:2024rsn} (see also Refs.~\cite{Alexandre:2024nuo, Thoss:2024hsr, Zantedeschi:2024ram, Boccia:2025hpm, Dondarini:2025ktz}). The white area represents the parameter space where memory-burdened PBHs can account for the total DM component of the Universe ($f_{\rm PBH} = 1$), whereas the hatched region indicates PBHs that have fully evaporated over cosmological times. Right: ratio of the limits on $M_{\rm PBH}$ placed by the different data samples over the global bound in order to highlight the most constraining measurement.}
    \label{fig:limit_k}
\end{figure*}

Fig.~\ref{fig:limit_k} shows the limit on $M_{\rm PBH}$ as a function of $k$, assuming $f_{\rm PBH} = 1$. The hatched region reports the region where PBHs have fully evaporated over cosmological timescales, while the white portion above the constraints highlights the region where PBHs can fully constitute DM. The lines correspond to the obtained limits, demonstrating that for $k > 3$, CRs provide constraints slightly stronger than UHE gamma-rays, complementing their information regarding PBH constraints. For $k \lesssim 1.4$, instead, UHECRs are not sensitive to the emission from memory-burdened PBHs, which appears at lower energies (see Fig.~\ref{fig:flux_proton}).

\section{Conclusions}
\label{sec:conclusions}

In this paper, we have explored the emission of UHECRs from memory-burdened PBHs. Indeed, the evaporation process induces a copious production of quarks which subsequently hadronize emitting protons and neutrons. Using the publicly-available \texttt{BlackHawk} and \texttt{HDMSPectra} codes, we have modeled the secondary proton and neutron emissions from a population of PBHs within the Milky Way, evaluating constraints on the PBH parameter space using limits imposed by the non-obervations of neutrons and protons at the highest energies by the Pierre Auger Observatory.

Despite implementing conservative statistical analyses, we have demonstrated that constraints with protons, for $k\gtrsim 3$ and $M_{\rm PBH} \lesssim 500\, \rm g$, are as tight as than previous ones obtained with UHE gamma-rays, despite those are not only emitted as secondary particles during hadronization but also as primary ones.

By contrast, limits obtained from neutrons are generally weaker than those derived from gamma rays, as they similarly rely on integral flux limits and associated with an overall lower flux. Nonetheless, neutron bounds remain competitive, reaching sensitivities comparable to those obtained with high-energy neutrinos~\cite{Chianese:2024rsn}. Future extension of this analysis might include studying the possibility of mergers of light memory-burden PBHs, which might lead to a new phase of evaporation, emitting high-energy particles~\cite{Dondarini:2025ktz,Zantedeschi:2024ram}. Moreover, future analyses could significantly improve these constraints by exploiting data by AugerPrime~\cite{Castellina:2019irv}--the upgrade of the Pierre Auger Experiment-- as well as by implementing dedicated spatial-template analyses, potentially leading to much tighter bounds.

\section*{Acknowledgements}

AA and CE gratefully acknowledge the PAO Collaboration for helpful discussions that contributed to this work. The authors also acknowledge Daniele Gaggero for fruitful discussions throughout several stages of the manuscript.
AA acknowledges the support of the project ``NUSES - A pathfinder for studying astrophysical neutrinos and electromagnetic signals of seismic origin from space'' (Cod.~id.~Ugov: NUSES; CUP: D12F19000040003). The work of CE has been partially funded by the European Union – Next Generation EU, through PRIN-MUR 2022TJW4EJ and by the European Union – NextGenerationEU under the MUR National Innovation Ecosystem grant ECS00000041 – VITALITY/ASTRA – CUP D13C21000430001. 
AA, MC and CE acknowledge the support by the research project TAsP (Theoretical Astroparticle Physics) funded by the Istituto Nazionale di Fisica Nucleare (INFN).

\bibliography{bibliography}
\end{document}